\documentclass[aps,pra,twocolumn,amsmath,amssymb,nofootinbib,superscriptaddress]{revtex4-1}
\usepackage{times}
\usepackage[pdftex]{graphicx}
\usepackage{dcolumn}
\usepackage{bm}
\usepackage{amsmath}
\usepackage{indentfirst}
\usepackage{float}
\usepackage[colorlinks]{hyperref}
\usepackage[dvipsnames]{xcolor}

\usepackage{makecell}

\newcommand{\sgz}{\hat{\sigma}^z}
\newcommand{\sgp}{\hat{\sigma}^+}
\newcommand{\sgm}{\hat{\sigma}^-}

\newcommand{\omegap}{\Omega_P}
\newcommand{\omegas}{\Omega_S}

\newcommand{\Hop}{\hat{H}}
\newcommand{\Fop}{\hat{F}}
\newcommand{\Fdop}{\hat{F}^{\dagger}}

\newcommand{\Dop}{\mathcal{D}}

\newcommand{\rhoop}{\hat{\rho}}
\newcommand{\im}{{\rm i}}

\begin{document}

\title{Population transfer under local dephasing}
\author{Wei Huang}
\affiliation{Guangxi Key Laboratory of Optoelectronic Information Processing, Guilin University of Electronic Technology, Guilin 541004, China}

% \author{Baohua Zhu}
% \affiliation{School of material science and engineering, Guilin University of Electronic Technology, Guilin 541004, China}

\author{Wentao Zhang}
\email{zhangwentao@guet.edu.cn}
\affiliation{Guangxi Key Laboratory of Optoelectronic Information Processing, Guilin University of Electronic Technology, Guilin 541004, China}

\author{Chu Guo}
\email{guochu604b@gmail.com}
\affiliation{Key Laboratory of Low-Dimensional Quantum Structures and Quantum Control of Ministry of Education, Department of Physics and Synergetic Innovation Center for Quantum Effects and Applications, Hunan Normal University, Changsha 410081, China}

\begin{abstract}
Stimulated Raman adiabatic passage is a well-known technique for quantum population transfer due to its robustness again various sources of noises. Here we consider quantum population transfer from one spin to another via an intermediate spin which subjects to dephasing noise. We obtain an analytic expression for the transfer efficiency under a specific driving protocol, showing that dephasing could reduce the transfer efficiency, but the effect of dephasing could also be suppressed with a stronger laser coupling or a longer laser duration. 
% An additional adiabatic condition is obtained on top of the one required by the standard STIRAP, which is related to the dephasing strength.
We also consider another commonly used driving protocol, which shows that this analytic picture is still qualitatively correct.
\end{abstract}

\date{\today}
\maketitle

\address{}

\vspace{8mm}

\section{Introduction}
Complete population transfer from an initial state to a final state has profound importance in both quantum and classical physics. On the quantum part, it has long been an active research area in quantum optics~\cite{Kuklinski1989,Bergmann1998,Huang2017}, and it is a fundamental technique for the physical realization of quantum information processing~\cite{Falci2017,Chen2018,KumarParaoanu2016,Helm2018,Chakraborty2017,Dory2016}. On the classical part, it has been used as a technique to achieve power or intensity inversion in classical systems~\cite{bergmann2019}, such as waveguide couplers~\cite{Huang2014}, wireless energy transfer~\cite{Rangelov2011}, and graphene systems~\cite{Huang2018sst,Huang2018carbon}.

Stimulated Raman adiabatic passage (STIRAP) has been one of the most important techniques for complete population transfer. In its standard implementation, an initial state $\vert 1\rangle$ and a final state $\vert 3\rangle$ are coupled to a common intermediate state $\vert 2\rangle$, by a pump laser and a Stocks laser respectively~\cite{GaubatzBergmann1990}. Complete population transfer between states $\vert 1\rangle$ and $\vert 3\rangle$ can then be achieved if the laser pulses are applied adiabatically and in a counter-intuitive order (the Stocks laser applies first), with the adiabatic condition
\begin{align}\label{eq:adia_cond}
\dot{\theta}(t) \ll \Omega(t).
\end{align}
Here $\Omega(t) = \sqrt{\omegap^2(t) + \omegas^2(t)}$ with $\omegap(t)$ and $\omegas(t)$ the Rabi frequencies of the pump laser and Stocks laser respectively, and $\tan(\theta(t)) = \omegap(t) / \Omega_S(t)$. STIRAP has  important advantages that it is robust against to the variations of the experimental conditions~\cite{VitanovBergmann2017}, and against to the decaying of the intermediate state~\cite{Glushko1992,Fleischhauer1996,VitanovStenholm1997}. STIRAP via multiple intermediate states has also been considered~\cite{VitanovStenholm1999}, as well as generalizations to intermediate state as a continuum~\cite{CarrollHioe1992,CarrollHioe1993,Nakajima1994,VitanovStenholm1997} and a lossy continuum~\cite{HuangGuo2019b}, where it is shown that significant partial population transfer can still be achieved. Recently, STIRAP via a thermal state or a thermal continuum has been studied, showing that the efficiency of population transfer will be reduced significantly in this case~\cite{HuangGuo2020}.

In this work, we focus on the effect of dephasing on the efficiency of population transfer via STIRAP. Dephasing is a common type of noise in quantum systems which induces decay of the off-diagonal terms in the density operator $\rhoop$. The standard three-level STIRAP with dephasing for all energy levels has already been considered in Ref.~\cite{IvanovBergmann2004}, where it is shown that the population of the final state $\rho_{33}(t)$ approximately satisfies
\begin{align}\label{eq:classical}
\rho_{33} = \frac{1}{3} + \frac{2}{3} e^{-\gamma_{13}\eta},
\end{align}
with $\eta = \frac{3}{4}\int_{-\infty}^{\infty}dt\sin^2\left(2\theta(t)\right)$. Here $\rho_{ij}$ denotes the element of $\langle i\vert\rhoop\vert j\rangle$ for $i, j \in \{1,2,3\}$, and $\gamma_{ij}$ denotes the decay rate of the element $\rho_{ij}$ for $i \neq j$. We note that in deriving Eq.(\ref{eq:classical}), the terms proportional to $\dot{\theta}(t)$ have been neglected and one is left with an expression which is independent of $\gamma_{12}$ or $\gamma_{23}$ or the laser strength $\Omega(t)$.

Here we consider a slightly different physical setup. Concretely, we study population transfer from a spin $q_1$ to another spin $q_3$, via an intermediate spin $q_2$. We assume that $q_2$ subjects to dephasing noise, while $q_1$ and $q_3$ are noise-free. The relevant states, namely $\{\vert 100\rangle, \vert 010\rangle, \vert 001\rangle\}$, form a three-level system which has a one-to-one correspondence with the standard three-level STIRAP.
A possible physical setup of our model is the information transfer between two well-protected cavities via a dephasing channel due to the decoherence. 
% ($\vert 100\rangle\rightarrow\vert 1\rangle$, $\vert 010\rangle \rightarrow \vert2\rangle$ and $\vert 001 \rangle \rightarrow \vert 3\rangle$).
 % with $\vert 100\rangle\rightarrow\vert 1\rangle$, $\vert 010\rangle \rightarrow \vert2\rangle$ and $\vert 001 \rangle \rightarrow \vert 3\rangle$. 
We derive an analytic expression for transfer efficiency under a specific driving protocol. Based on this expression we then obtain an additional adiabatic condition on top of Eq.(\ref{eq:adia_cond}), which is related to the dephasing strength. We show that dephasing reduces the transfer efficiency. However, the effect of dephasing could be suppressed by a stronger laser coupling or a longer laser duration. The paper is organized as follows. We introduce our model in Sec.\ref{sec:spin}. Then in Sec.\ref{sec:results}, we derive the analytic expression for the transfer efficiency as well as the additional adiabatic condition under which complete population transfer can still be achieved. We show that the analytic expression agrees well with the predictions from the exact quantum master equation in a wide parameter range with numerical simulations. We conclude in Sec.\ref{sec:summary}.

% STIRAP via a dephasing state or a dephasing continuum. 

\section{Model}\label{sec:spin}

\begin{figure}
\includegraphics[width=\columnwidth]{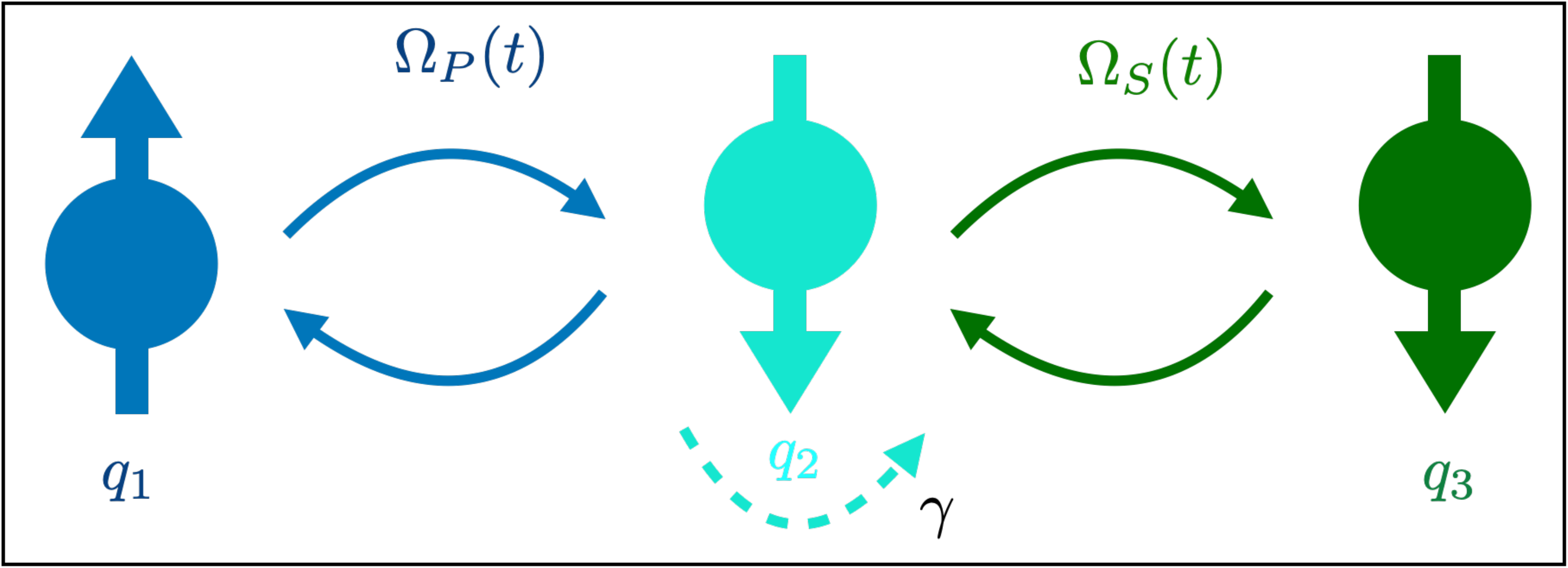}
\caption{Population transfer from spin $q_1$ to $q_3$ via an intermediate spin $q_2$ which subjects to dephasing noise with strength $\gamma$. The pump laser $\omegap(t)$ couplings $q_1$ and $q_2$ while the Stocks laser $\omegas(t)$ couples $q_2$ and $q_3$. The initial state $\vert q_1 q_2 q_3\rangle$ of the system is chosen as $\vert 100\rangle$ and the final state would be $\vert 001\rangle$ if complete population transfer is achieved.}
\label{fig:fig1}
\end{figure}

Our model consists of three spins in which the intermediate spin acts as a bus for population transfer and subjects to dephasing, which is shown in Fig.~\ref{fig:fig1}. The quantum Lindblad master equation describes the equation of motion~\cite{Lindblad1976,Gorini1976}, which is
\begin{align}\label{eq:lindblad}
\frac{d\rhoop(t)}{dt} = -\im [\Hop(t), \rhoop] + \Dop(\rhoop),
\end{align}
where the Hamiltonian $\Hop(t)$ takes the form
\begin{align}\label{eq:ham}
\Hop(t) =& \Delta \sum_{j=1}^3\sgz_j + \omegap(t)\left(\sgp_1\sgm_2 + \sgm_1\sgp_2\right) \nonumber \\
&+ \omegas(t)\left(\sgp_2\sgm_3 + \sgm_2\sgp_3\right),
\end{align}
with $\Delta$ the energy difference for all spins. The pump laser $\omegap(t)$ couples the $1$-th spin to the $2$-th spin, while the Stocks laser couples the $3$-th spin to the $2$-th spin. The dissipator $\Dop$ takes the form
\begin{align}
\Dop(\rhoop) = \gamma \left(\sgz_2 \rhoop \sgz_2 - \rhoop \right),
\end{align}
with $\gamma$ the dephasing strength. The Hamiltonian as well as the dissipation conserves the total number of excitations. Since the initial state in the context of STIRAP is chosen as $\vert 100\rangle$, we are restricted to a subspace spanned by three states $\{\vert 100\rangle, \vert 010\rangle, \vert 001\rangle\}$ only. We can see that our model remains the same if the intermediate spin is replaced by a dephasing bosonic mode and rotating wave approximation is applied to the spin-boson couplings, where the intermediate bosonic mode may be physically implemented using a cavity as the "flying qubit". It is also possible to generalize the intermediate state in our model as a chain of spins as in straddle STIRAP~\cite{Vitanov19981,Vitanov19982}, or a bosonic continuum~\cite{HuangGuo2019b,HuangGuo2020} under dephasing.

Now using the mapping 
\begin{align}\label{eq:mapping}
\vert 100\rangle \leftrightarrow \vert 1\rangle, \vert 010\rangle \leftrightarrow \vert2\rangle, \vert 001 \rangle \leftrightarrow \vert 3\rangle,
\end{align}
the Hamiltonian in Eq.(\ref{eq:ham}) can be rewritten in the basis of $\{\vert 1\rangle, \vert 2\rangle, \vert 3\rangle\}$ as
\begin{align}\label{eq:h3}
\Hop(t) = \left[\begin{array}{ccc}
0 & \omegap(t) & 0 \\
\omegap(t) & 0 & \omegas(t) \\
0 & \omegas(t) & 0
\end{array}\right],
\end{align}
and the dissipator $\Dop$ can be written as 
\begin{align}\label{eq:dop}
\Dop(\rhoop) = \gamma\left(\Fop \rhoop \Fdop - \rhoop\right) = -2\gamma \left[ \begin{array}{ccc}
0 & \rho_{12} & 0 \\
\rho_{21} & 0 & \rho_{23} \\
0 & \rho_{32} & 0 
\end{array} \right],
\end{align}
with 
\begin{align}
\Fop = \left[\begin{array}{ccc}
-1 & 0 & 0 \\
0 & 1 & 0 \\
0 & 0 & -1 
\end{array}\right]
\end{align}
being the operator $\sgz_2$ in the new basis. Compared to the model studied in Ref.~\cite{IvanovBergmann2004}, we can see from Eq.(\ref{eq:dop}) that we have $\gamma_{13}=0$. In this case Eq.(\ref{eq:classical}) predicts complete population transfer irrespective of the value of $\gamma$. However as will be clear later, $\gamma$ would  significantly suppress the transfer efficiency in this case if it is comparable to other parameters such as $\Omega(t)$. Therefore in the following we perform a more refined calculation for the population transfer efficiency which would allow us the see more clearly the role of dephasing.

\section{Results and discussions}\label{sec:results}

The Hamiltonian in Eq.(\ref{eq:h3}) can be diagonalized with three instaneous eigenstates
\begin{align}
\vert +\rangle =&  \frac{\sqrt{2}}{2} \left(\sin(\theta)  \vert 1\rangle + \vert 2\rangle + \cos(\theta)\vert 3\rangle \right); \\
\vert d\rangle = & \cos(\theta) \vert 1\rangle - \sin(\theta)\vert 3\rangle; \label{eq:dark} \\
\vert -\rangle =& \frac{\sqrt{2}}{2} \left(\sin(\theta)  \vert 1\rangle - \vert 2\rangle + \cos(\theta)\vert 3\rangle\right), 
\end{align}
corresponding to eigenvalues $\Omega(t)$, $0$, and $-\Omega(t)$ respectively. 
In the ideal three-level STIRAP, one adiabatically changes $\theta(t)$ from $0$ to $\pi/2$ by tuning the ratio $\omegap(t) / \omegas(t)$ such that once the initial state is chosen to be $\vert 1\rangle$, it will always remain in $\vert d\rangle$. The unitary matrix $W$ to diagonalize $\Hop(t)$ can be written as
\begin{align}\label{eq:w}
W(\theta) = \left[ \begin{array}{ccc}
\frac{\sqrt{2}}{2} \sin(\theta) & \cos(\theta) & \frac{\sqrt{2}}{2}\sin(\theta) \\
\frac{\sqrt{2}}{2} & 0 & -\frac{\sqrt{2}}{2} \\
\frac{\sqrt{2}}{2} \cos(\theta) & -\sin(\theta) & \frac{\sqrt{2}}{2}\cos(\theta)
\end{array} \right].
\end{align}
To gain better insight into the time evolution, we go to the adiabatic picture, in which the Eq.(\ref{eq:lindblad}) becomes~\cite{IvanovBergmann2004}
\begin{align}\label{eq:lindblada}
\frac{d\rhoop_a}{dt} = -\im [\Hop_a, \rhoop_a] - [M, \rhoop_a] + \Dop_a(\rhoop_a).
\end{align}
Here $\rhoop_a$, $\Hop_a$, $\Dop_a$ are the corresponding operators to $\rhoop$, $\Hop$, $\Dop$ respectively in the adiabatic basis $\{\vert +\rangle, \vert d\rangle, \vert -\rangle\}$, which can be written as
\begin{align}
\rhoop_a =& W^{\dagger}\rhoop W = \left[\begin{array}{ccc}
\rho_{++} & \rho_{+d} & \rho_{+-} \\
\rho_{d+} & \rho_{dd} & \rho_{d-} \\
\rho_{-+} & \rho_{-d} & \rho_{--}
\end{array}\right]; \label{eq:rhoa} \\
\Hop_a(t) =& W^{\dagger} \Hop W = \Omega(t) \left[\begin{array}{ccc}
1 & 0 & 0 \\
0 & 0 & 0 \\
0 & 0 & -1
\end{array}\right]; \label{eq:hopa} \\
\Dop_a(\rhoop_a) =& \gamma\left(\Fop_a \rhoop_a \Fdop_a - \rhoop_a\right), \label{eq:dopa} 
\end{align}
with 
\begin{align}
\Fop_a =& W^{\dagger} \Fop W =  \left[\begin{array}{ccc}
0 & 0 & -1 \\
0 & -1 & 0 \\
-1 & 0 & 0
\end{array}\right].
\end{align}
Here we have used $\rho_{uv}$ with $u, v \in \{+, d, -\}$ to denote the element $\langle u\vert \rhoop_a\vert v\rangle$.
The gauge matrix $M$ satisfies
\begin{align}\label{eq:ma}
M = W^{\dagger} \dot{W} = \frac{\sqrt{2}}{2} \dot{\theta} \left[\begin{array}{ccc}
0 & -1 & 0 \\
1 & 0 & 1 \\
0 & -1 & 0
\end{array}\right],
\end{align}
which results from the time dependence of the adiabatic basis. Now substituting Eqs.(\ref{eq:hopa}, \ref{eq:dopa}, \ref{eq:ma}) into Eq.(\ref{eq:lindblada}), we get the following set of equations
\begin{subequations}\label{eq:ana6}
\begin{align}
\dot{\rho}_{++} =& \frac{\sqrt{2}}{2}\dot{\theta}\left( \rho_{d+} + \rho_{+d} \right) + \gamma\left(\rho_{--} - \rho_{++}\right); \label{eq:rhopp} \\
\dot{\rho}_{--} =& \frac{\sqrt{2}}{2}\dot{\theta}\left( \rho_{d-} + \rho_{-d} \right) + \gamma\left(\rho_{++} - \rho_{--}\right); \label{eq:rhomm} \\
\dot{\rho}_{dd} =& -\frac{\sqrt{2}}{2}\dot{\theta}\left(\rho_{+d} +\rho_{d+} + \rho_{d-} +\rho_{-d}\right); \label{eq:rho00} \\
\dot{\rho}_{+d} =& -\im \Omega\rho_{+d} -\frac{\sqrt{2}}{2}\dot{\theta}\left(-\rho_{dd} + \rho_{++}+\rho_{+-}\right) \nonumber \\ 
&+ \gamma\left(\rho_{-d} - \rho_{+d}\right); \label{eq:rhop0} \\
\dot{\rho}_{d-} =& -\im \Omega\rho_{d-}-\frac{\sqrt{2}}{2}\dot{\theta}\left(- \rho_{dd} + \rho_{+-} + \rho_{--}\right) \nonumber \\ 
&+ \gamma\left(\rho_{d+} - \rho_{d-}\right); \label{eq:rho0m} \\
\dot{\rho}_{+-} =& -2\im \Omega\rho_{+-}+\frac{\sqrt{2}}{2}\dot{\theta}\left(\rho_{d-}+\rho_{+d}\right) \nonumber \\ 
&+ \gamma \left(\rho_{-+} - \rho_{+-}\right), \label{eq:rhopm}
\end{align}
\end{subequations}
We note that in Ref.~\cite{IvanovBergmann2004}, the second term in Eq.(\ref{eq:lindblada}) is neglected since it depends on $\dot{\theta}$ which is assumed to be small. However in our case if this term is neglected, we will arrive at a solution where the population is trapped in $\vert d\rangle$, since it is a dark state of the dissipator $\Dop_a$.

The set of Eqs.(\ref{eq:ana6}) are difficult to solve analytically in general. However, they can be significantly simplified with several reasonable assumptions. First, in the context of STIRAP, the adiabatic condition in Eq.(\ref{eq:adia_cond}) requires $\dot{\theta}$ to be smaller compared to other relevant parameters. Second, we assume that in the adiabatic basis the off-diagonal terms of $\rhoop_a$ are small, that is $\rho_{uv} \ll 1$ if $u \neq v$. Now we subtract Eq.(\ref{eq:rhomm}) from Eq.(\ref{eq:rhopp}) and get an equation for $g = \rho_{++} - \rho_{--} $ as
\begin{align}\label{eq:g}
\dot{g} = \frac{\sqrt{2}}{2} \dot{\theta} \left(\rho_{d+} + \rho_{+d} - \rho_{d-} - \rho_{-d} \right) - 2\gamma g.
\end{align}
The first term on the right-hand side of Eq.(\ref{eq:g}) contains $\dot{\theta}$ and $\rho_{d+} + \rho_{+d} - \rho_{d-} - \rho_{-d}$ which are both small numbers. Thus we neglect this term and get $\dot{g} = - 2\gamma g$. Since $g(t)$ is initially $0$, and get $g(t)=0$ for all $t$, namely 
\begin{align}\label{eq:s1}
\rho_{++}(t) = \rho_{--}(t).
\end{align}
Similarly, subtracting Eq.(\ref{eq:rho0m}) from Eq.(\ref{eq:rhop0}), we get an equation for $h = \rho_{+d} - \rho_{d-}$ as
\begin{align}\label{eq:h}
\dot{h} = -\im \Omega h - \gamma(h + h^{\ast}),
\end{align}
where we have used $\rho_{++}(t) = \rho_{--}(t)$. Now since $h(t)$ is initially $0$, from Eq.(\ref{eq:h}) we have $h(t)=0$ for all $t$, namely 
\begin{align}\label{eq:s2}
\rho_{+d}(t) = \rho_{d-}(t).
\end{align}
Finally, the second term on the right-hand side of Eq.(\ref{eq:rhopm}) can be neglected for the same reason, and we get
\begin{align}\label{eq:rhopm2}
\dot{\rho}_{+-} = -2\im \Omega\rho_{+-} + \gamma \left(\rho_{-+} - \rho_{+-}\right).
\end{align}
Since $\rho_{+-}$ is initially $0$, from Eq.(\ref{eq:rhopm2}) we get 
\begin{align}\label{eq:s3}
\rho_{+-}(t)=0
\end{align}
for all $t$. Substituting Eqs.(\ref{eq:s1}, \ref{eq:s2}, \ref{eq:s3}) back into Eqs.(\ref{eq:ana6}), and assuming $\rho_{+d} = a + \im b$ with $a(t)$ and $b(t)$ real functions, we get the following closed set of equations for $\rho_{dd}$, $a$, $b$
\begin{subequations}\label{eq:ana3}
\begin{align}
\dot{\rho}_{dd} =& -2\sqrt{2}\dot{\theta} a; \label{eq:ana3a} \\
\dot{a} =& \Omega b - \frac{\sqrt{2}}{4}\dot{\theta}\left(1-3\rho_{dd}\right); \\
\dot{b} = & -\Omega a - 2\gamma b .
\end{align}
\end{subequations}
Eqs.(\ref{eq:ana3}) are still difficult to solve analytically since their coefficients are time-dependent in the general case. 

\begin{figure}
\includegraphics[width=\columnwidth]{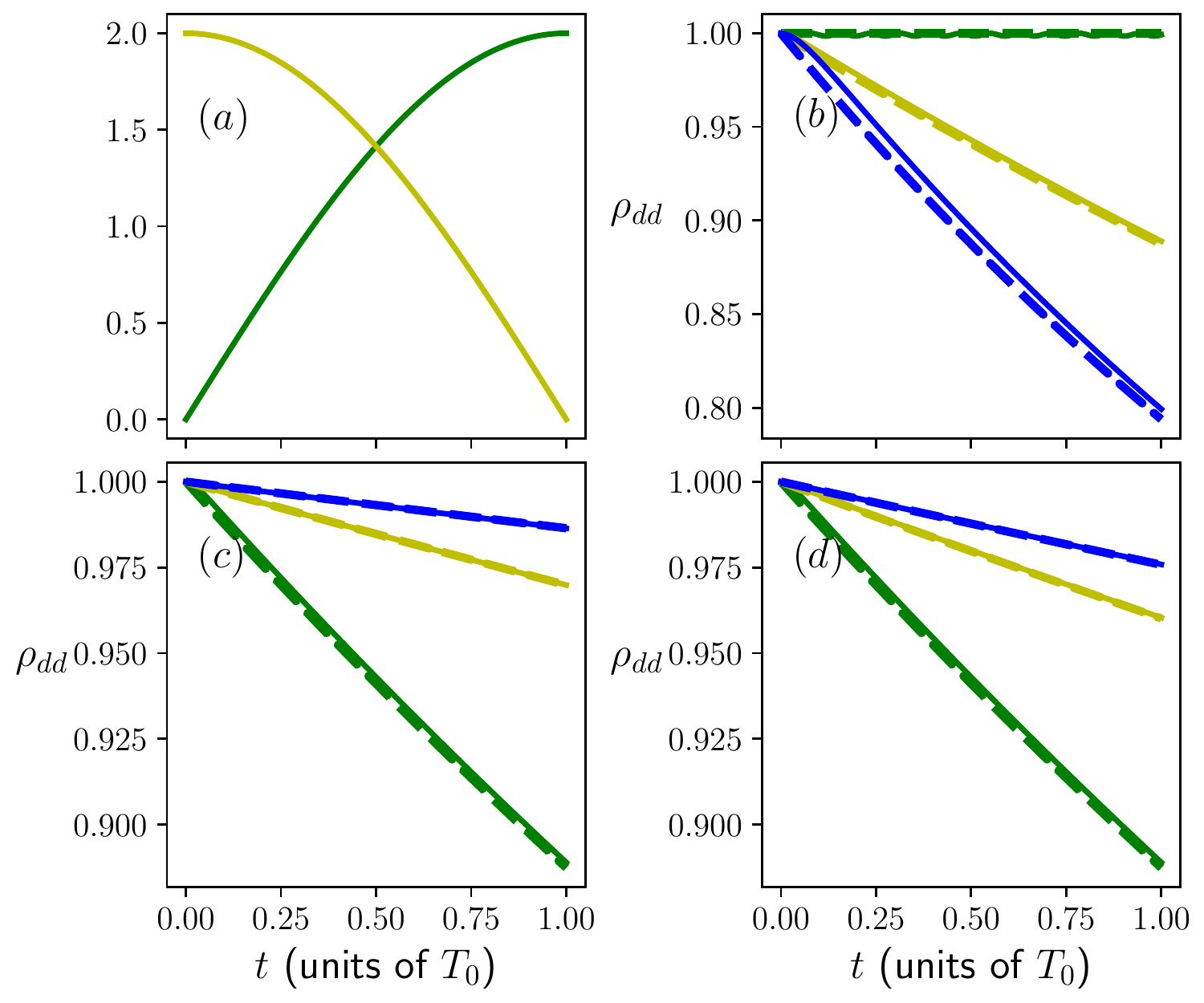}
\caption{(a) The driving protocol as in Eq.(\ref{eq:protocolb}) at $\Omega_0=2$ and $T_0=40$, the yellow and green solid lines represent $\omegap(t)$ and $\omegas(t)$ respectively. (b) The green, yellow, blue lines from top down show the final occupation on the dark state $\vert d\rangle$, $\rho_{dd}$, as a function of time $t$ for $\gamma=0,2,4$ respectively. (c) The green, yellow, blue lines from down to top show $\rho_{dd}$ as a function of time $t$ for $\Omega_0=2,4,6$, respectively. (d) The green, yellow, blue lines from down to top show $\rho_{dd}$ as a function of time $t$ for $T_0=40,120,200$ respectively. In (b,c,d) the solid and dashed lines represent the exact numerical solutions from Eq.(\ref{eq:lindblada}) and the analytic solutions from Eq.(\ref{eq:ana1}), respectively. The other parameters used are $\gamma=2$, $\Omega_0=2$, $T_0=40$ if not particularly specified.}
\label{fig:fig2}
\end{figure}

\begin{figure}
\includegraphics[width=\columnwidth]{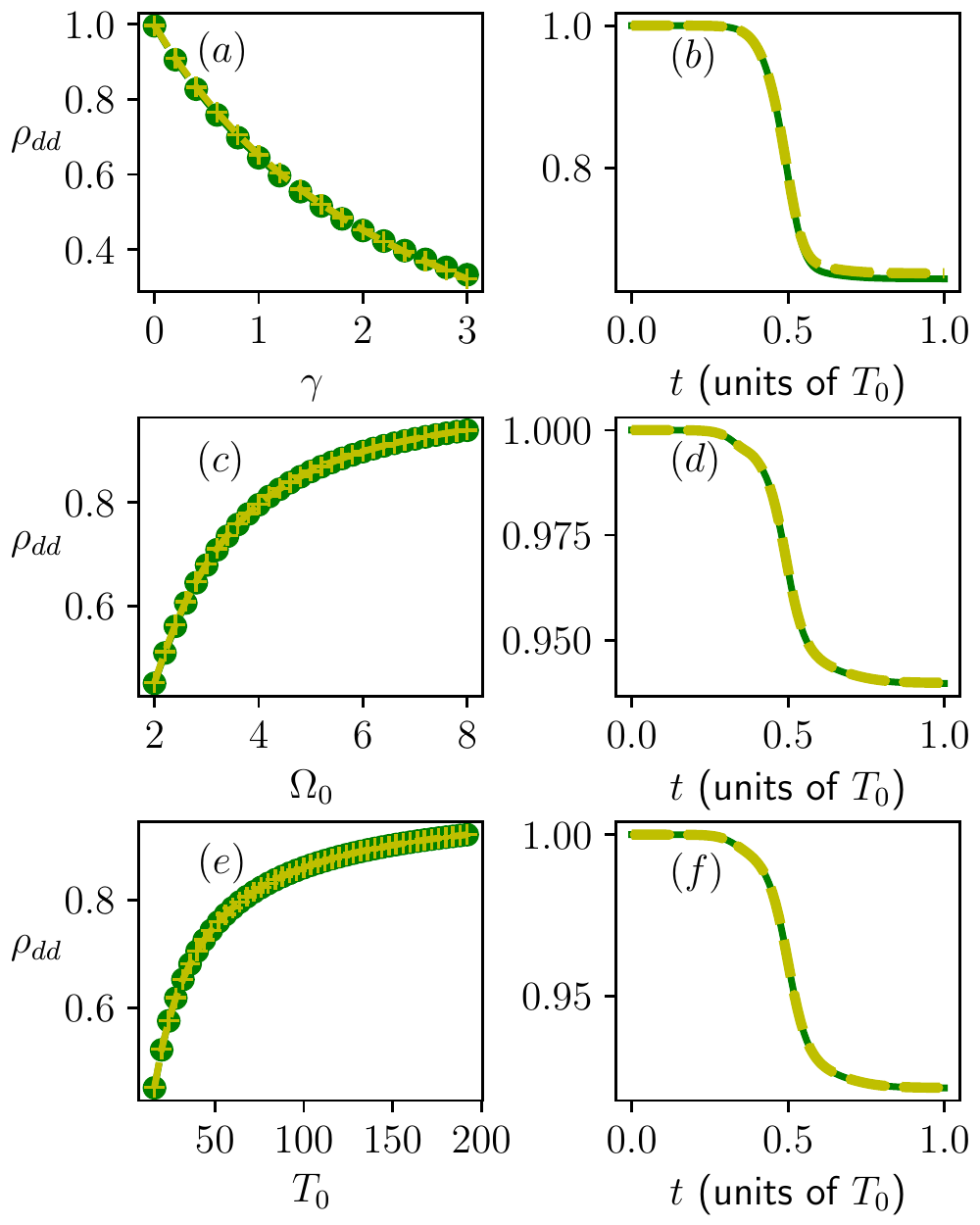}
\caption{(a,c,e) The final occupation on the dark state $\vert d\rangle$, $\rho_{dd}$, as a function of $\gamma$, $\Omega_0$ and $T_0$ respectively. The green dashed line with circle represents the exact solutions of Eq.(\ref{eq:lindblada}), while the yellow dashed line with $+$ represents the analytic solutions of Eqs.(\ref{eq:ana3}). (b, d, f) $\rho_{dd}$ as a function of $t$ for $\gamma=1$, for $\Omega_0 = 8$, and for $T_0=192$ respectively. The green solid line and the yellow dashed line stand for the exact solutions of Eq.(\ref{eq:lindblada}) and analytic solutions of Eq.(\ref{eq:ana3}) respectively. The other parameters used are $\tau=1$, $\gamma=2$, $\Omega_0 = 2$, $T_0 = 16$ if not particularly specified. }
\label{fig:fig3}
\end{figure}

In the following, we consider a specific driving protocol as follows:
\begin{subequations}\label{eq:protocolb}
\begin{align}
\omegap(t) = \Omega_0 \sin(\frac{\pi t}{2 T_0}); \\
\omegas(t) = \Omega_0 \cos(\frac{\pi t}{2 T_0}),
\end{align}
\end{subequations}
where $\Omega_0$ denotes the strength of the laser coupling and $T_0$ is the total duration of it. We can see that $\omegap(0)/\omegas(0) = 0$ and $\omegap(T_0)/\omegas(T_0) = \infty$.
The advantage of the protocol in Eq.(\ref{eq:protocolb}) is that we have $\Omega(t) = \Omega_0$, $\theta(t) = \frac{\pi t}{2T_0}$ and $\dot{\theta} = \frac{\pi}{2T_0}$. We further assume that in Eqs.(\ref{eq:ana3}) $a(t)$ and $b(t)$ are slowly varying variables compared to $\rho_{00}(t)$. As a result we can set $\dot{a}=\dot{b}=0$ and then Eqs.(\ref{eq:ana3a}) can be solved as
\begin{align}\label{eq:ana1}
\rho_{dd}(t) = \frac{1}{3} + \frac{2}{3}e^{-3 \chi t},
\end{align}
with 
\begin{align}\label{eq:chi}
\chi = \frac{2\gamma\dot{\theta}^2}{\Omega^2}.
\end{align}
To check the validity of Eq.(\ref{eq:ana1}), we compared it with the exact numerical solutions from Eq.(\ref{eq:lindblada}). In Fig.\ref{fig:fig3}(a) we show an instance of the driving protocol in Eq.(\ref{eq:protocolb}). In Fig.~\ref{fig:fig3}(b, c, d) we compare $\rho_{dd}$ predicted by Eq.(\ref{eq:ana1}) and by Eq.(\ref{eq:lindblada}) as functions of time $t$ versus different values of $\gamma$, $\Omega_0$, $T_0$ respectively. We can see that our analytic prediction agrees very well with the exact solution in a wide parameter range we have considered.  

To this end we discuss the implications of our analytic solution in Eq.(\ref{eq:ana1}). From Eq.(\ref{eq:dark}) we have $\rho_{33}(T_0) = \rho_{dd}(T_0)$. Therefore the final occupation of $\rho_{dd}(T_0)$ represents the population transfer efficiency. Then from Eq.(\ref{eq:ana1}) we have
\begin{align}\label{eq:rhof}
\rho_{33}(T_0) = \frac{1}{3} + \frac{2}{3}e^{-\frac{3\pi\gamma\dot{\theta}}{\Omega^2}},
\end{align}
where we have used $\dot{\theta} T_0 = \pi/2$. We can see that $\rho_{33}(T_0)$ decreases exponentially with $\gamma$. However, the effect of dephasing can be made arbitrarily small if we increase laser coupling strength $\Omega$ or increase the laser duration (such that $\dot{\theta}$ will be smaller). Interestingly, based on Eq.(\ref{eq:rhof}) we can define an additional adiabatic condition on top of Eq.(\ref{eq:adia_cond}) which takes the dephasing strength into account. The additional adiabatic condition is simply $-3\chi T_0 \ll 1$, which is
\begin{align}\label{eq:adia_cond2}
\dot{\theta} \ll \frac{\Omega^2}{3\pi\gamma}.
\end{align}
Complete population transfer can still be achieved as long as Eqs.(\ref{eq:adia_cond}, \ref{eq:adia_cond2}) are both satisfied. Now for comparison, we have $\eta = 3T_0/8$ in Eq.(\ref{eq:classical}) under the driving protocol in Eq.(\ref{eq:protocolb}), that is, the population transfer efficiency is independent of $\Omega$, but decreases exponentially both with $\gamma_{13}$ and $T_0$. Therefore complete population transfer can never be achieved as long as $\gamma_{13} \neq 0$, since we can neither tune $\dot{\theta}$ to be very small ($T_0$ will be very larger) or very large (which breaks the adiabatic condition in Eq.(\ref{eq:adia_cond})).

% obtain a more refined adiabatic condition on top of Eq.(\ref{eq:adia_cond}) in this case by taking into account dephasing, which is $-3\chi T_0 \ll 1$, which can be reduced to
% \begin{align}\label{eq:adia_cond2}
% \dot{\theta} \ll \frac{\Omega^2}{3\pi\gamma},
% \end{align}
% where we have used the fact $\dot{\theta} T_0 = \pi/2$. 

% Eq.(\ref{eq:classical}) predict a final occupation on the state $\vert 3\rangle$ which is $\frac{1}{3} + \frac{2}{3}e^{-\frac{3\gamma_{13}T_0}{8}}$.
% \begin{align}
% \rho^{\prime}_{33}(T_0) = \frac{1}{3} + \frac{2}{3}e^{-\frac{3\gamma_{13}T_0}{8}}.
% \end{align}

% From Eqs.(\ref{eq:ana1}, \ref{eq:chi}) we can see that the negative effect of $\gamma$ can be suppressed by either increasing $\Omega$ or increasing $T$, this agree with the numerical results in Fig.~\ref{fig:fig2} and Fig.~\ref{fig:fig3}. 

Our analytic solution in Eq.(\ref{eq:ana1}) does not hold for general laser driving protocols. To show the validity of the physical picture we obtained based on the specific driving protocol in Eq.(\ref{eq:protocolb}), we numerically study the effect of local dephasing under the commonly used Gaussian driving protocol as follows
\begin{align}\label{eq:protocola}
\omegap(t) =& \Omega_0 \exp\left(-\frac{\left(t - \tau/2 - T_0/2 \right)^2}{T^2}\right); \\
\omegas(t) =& \Omega_0 \exp\left(-\frac{\left(t + \tau/2 - T_0/2 \right)^2}{T^2}\right).
\end{align}
Here $T$ denotes the width of the Gaussian laser coupling, $\tau$ is the delay between the two lasers, $T_0$ is the duration of the lasers which we choose as $T_0 = 8T$.
The population $\rho_{dd}$ on the dark state $\vert d\rangle$ as functions of $\gamma$, $\Omega_0$ and $T_0$, are shown in Fig.~\ref{fig:fig2}, where we have also checked the validity of Eqs.(\ref{eq:ana3}) by comparing its solutions to the exact solutions from Eq.(\ref{eq:lindblada}). From Fig.~\ref{fig:fig2}(a), we can see that dephasing can significantly suppress population transfer. While from Fig.\ref{fig:fig2}(c, e), we can see that significant population transfer can be restored by increasing $\Omega_0$ or $T_0$. This demonstrates that the physical picture obtained from our analytic solution is still valid for other laser driving protocols. Additionally, we can see that the simplified set of equations as in Eqs.(\ref{eq:ana3}) agree very well with the exact Lindblad equation in the wide parameter range considered in Fig.~\ref{fig:fig3}.

\section{Conclusion}\label{sec:summary}
To summarize, we have considered population transfer using STIRAP between two spins via an intermediate spin which subjects to dephasing, while these two spins themselves are dephasing-free. We derive an analytic expression for the population transfer efficiency under a specific laser driving protocol. Based on the analytical expression, we obtain an additional adiabatic condition which is related to the strength of dephasing, under which complete population transfer could still be achieved. We show that population transfer efficiency would be reduced by dephasing, but could be restored by using a stronger laser coupling or a longer laser duration. We have also shown that this physical picture is still qualitatively correct for the commonly used Gaussian type of laser driving. Our result is helpful for a better understanding the effect of dephasing on the quantum population transfer based on STIRAP.

% be alleviated by either a stronger laser coupling a longer laser duration. For the case of single intermediate spin, we derive analytic expression for the population transfer efficiency with a particular driving protocol, which allows us to quantitively understand the role of dephasing and give an additional adiabatic condition involving the strength of dephasing.

\begin{acknowledgments}
This work is acknowledged for funding National Science and Technology Major Project (2017ZX02101007-003), National Natural Science Foundation of China  (61965005). W.H. is acknowledged for funding from Guangxi oversea 100 talent project and W.Z. is acknowledged for funding from Guangxi distinguished expert project. C. G acknowledges support from National Natural Science Foundation of China under Grant No. 11805279.
\end{acknowledgments}

\bibliographystyle{apsrev4-1}
\bibliography{refs}

\end{document}